\documentclass[10pt]{article}
\usepackage{charter} 
\usepackage{fullpage}
\usepackage{cite}
\usepackage{amsmath,amssymb,amsfonts}
\usepackage{graphicx}
\usepackage{textcomp}

\usepackage{xr-hyper}
\usepackage[switch]{lineno}

\usepackage{graphics,amssymb,amsmath,epsfig,color}
\usepackage{graphicx}
\usepackage{multibib}
\usepackage{xspace}

\usepackage[hyphens]{url}     
\usepackage[hidelinks]{hyperref}       
\usepackage{booktabs}       
\usepackage{amsfonts}       
\usepackage{amssymb}        
\usepackage{nicefrac}       
\usepackage{microtype}      
\usepackage{bm}				
\usepackage{cite}			
\usepackage{graphicx}		
\usepackage{mathtools}		
\usepackage{algorithm}
\usepackage{algpseudocode}
\usepackage{amsthm}			
\usepackage{enumitem}		
\usepackage{multirow}       
\usepackage{tabularx}	    
\usepackage{hhline}
\usepackage{arydshln}		
\usepackage{xcolor}
\usepackage{mathtools}

\usepackage{booktabs}
\usepackage{multirow}       
\usepackage{tabularx}       
\usepackage{hhline}
\usepackage{arydshln}		
\usepackage{xcolor}
\usepackage[flushleft]{threeparttable}

\definecolor{lightgreen}{rgb}{.9,1,.9}

\newcolumntype{L}[1]{>{\raggedright\arraybackslash}p{#1}}
\newcolumntype{C}[1]{>{\centering\arraybackslash}p{#1}}
\newcolumntype{R}[1]{>{\raggedleft\arraybackslash}p{#1}}
\theoremstyle{plain} 

\def\defn{\,\coloneqq\,}
\def\argmin{\mathop{\mathrm{arg\,min}}} 

\def\lim{\mathop{\mathrm{lim}}} 

\def\log{\mathrm{log}}

\newcommand{\norm}[1]{\left\lVert#1\right\rVert}

\newcommand{\inmath}{\ensuremath}
\newcommand{\xmath}[1] {\inmath{#1}\xspace}
\newcommand{\blmath}[1] {\xmath{\bm{#1}}}

\newcommand{\x}{\blmath{x}}
\newcommand{\y}{\blmath{y}}
\newcommand{\D}{\blmath{D}}

\newcommand{\Dblur} {\xmath{\D_{\mathrm{blur}}}}
\newcommand{\Dscat} {\xmath{\D_{\mathrm{s}}}}
\newcommand{\Bscat} {\xmath{\bm{B}_{\mathrm{s}}}}

\newcommand{\Gblur} {\xmath{\bm{G}_{\mathrm{blur}}}}
\newcommand{\Gscat} {\xmath{\bm{G}_{\mathrm{scatter}}}}
\newcommand{\sblur} {\xmath{\sigma_{\mathrm{blur}}}}
\newcommand{\sscat} {\xmath{\sigma_{\mathrm{scatter}}}}

\def\ebm{{\bm{e}}}

\def\xbm{{\bm{x}}}

\def\ybm{{\bm{y}}}

\def\thetabm{{\bm{\theta}}}

\def\Abm{{\bm{A}}}

\def\Dbm{{\bm{D}}}

\def\Ibm{{\bm{I}}}

\def\xbmhat{{\widehat{\bm{x}}}}

\def\Rsf{{\mathrm{R}}}

\def\R{\mathbb{R}}

\def\Rsf{{\mathsf{R}}}

\def\thetabm{{\bm{\theta }}}

\def\df{g}  
\def\reg{r}

\long\def\red#1{\bgroup\color{red}#1\egroup}
\newcommand{\todo}[1]{{\color{black}#1}}

\begin{document}
	\title{Swap-Net: A Memory-Efficient 2.5D  Network
		\\
		for Sparse-View 3D Cone Beam CT Reconstruction}
	
	
	\author{
		Xiaojian Xu\thanks{X.~Xu, J.~Hu, and J.~A.~Fessler are with the 
			Department of Electrical Engineering \& Computer Science, University of Michigan, MI 48109, USA.} ,
		Marc~Klasky\footnotemark[2] ,
		Michael~T.~McCann\footnotemark[2]\thanks{M.~T.~McCann and M.~Klasky are with Theoretical Division, Los Alamos National Laboratory, Los Alamos, NM 87545, USA.} ,
		Jason~Hu\footnotemark[1] ,
		and Jeffrey~A.~Fessler\footnotemark[1] $^,$  \thanks{This material is based upon work supported by the Laboratory Directed Research and Development program of Los Alamos National Laboratory.}
		\vspace{-2\baselineskip} 
	}
	
	\date{}
	\maketitle
	\begin{abstract}
		Reconstructing 3D cone beam computed tomography (CBCT) images
		from a limited set of projections
		is an important inverse problem in many imaging applications
		from medicine to inertial confinement fusion (ICF).
		The performance of traditional methods such as filtered back projection (FBP)
		and model-based regularization is sub-optimal when the number of available projections is limited.
		In the past decade, deep learning (DL) has gained great popularity for solving CT inverse problems.
		A typical DL-based method for CBCT image reconstruction is to learn an end-to-end mapping
		by training a 2D or 3D network.
		However, 2D networks fail to fully use global information.
		While 3D networks are desirable, they become impractical
		as image sizes increase
		because of the high memory cost.
		This paper proposes Swap-Net, a memory-efficient 2.5D network
		for sparse-view 3D CBCT image reconstruction.
		Swap-Net uses a sequence of novel axes-swapping operations
		to produce 3D volume reconstruction in an end-to-end fashion without using full 3D convolutions.
		Simulation results show that Swap-Net consistently outperforms baseline methods
		both quantitatively and qualitatively in terms of reducing artifacts
		and preserving details of complex hydrodynamic simulations of relevance to the ICF community.
	\end{abstract}
	
	


	\section{Introduction}
	
	The recovery of high-quality images from limited projection measurements
	is fundamental in computed tomography (CT)~\cite{selig:24:ldc}.
	Cone beam CT (CBCT) is a specialized imaging technique used
	in fields requiring detailed 3D imaging.
	In CBCT, an X-ray beam is projected through the 3D object onto a 2D detector.
	Unlike traditional CT scanners where the X-ray beam is collimated into a narrow fan shape,
	CBCT systems use a cone-shaped beam,
	allowing wider coverage of the object in a single rotation.
	CBCT is a valuable tool in various applications
	for obtaining detailed structural information%
	~\cite{alamri2012applications, horner2015guidelines, casselman2013cone}.
	
	A CBCT scanner captions
	2D X-ray projections, also called radiographs,
	as it rotates around the target object.
	Computer algorithms process these projections
	to reconstruct a 3D volumetric image of the object.
	Developing fast and accurate methods for 3D CBCT image reconstruction
	is important in many applications~\cite{alamri2012applications, horner2015guidelines, casselman2013cone}.
	Filtered back projection (FBP) is a classical algorithm
	that is computationally efficient and relatively straightforward to implement
	\cite{feldkamp:84:pcb, pan2009commercial}.
	However, FBP is sensitive to measurement noise
	and leads to artifacts
	when given incomplete or irregularly sampled projection data.
	Regularized inversion methods view CT imaging as an \emph{inverse problem},
	where the unknown object is reconstructed by combining a CT physical model
	and a hand-crafted regularizer%
	~\cite{sauer1993local, kim2014sparse, fessler2000statistical, elbakri2002statistical, thibault2006recursive, beister2012iterative, zhang2016low, yu2017image, xu2020sparse}.
	Recently, \emph{deep learning (DL)} methods
	have gained popularity
	in solving CBCT inverse problems%
	~\cite{anirudh2018lose, ravishankar2019image, zheng2018pwls, chen2018learn, gupta2018cnn}.
	Traditional DL methods are based on training \emph{convolutional neural networks} (CNNs)
	to map the measurements or low-quality images to the desired high-quality images.
	\emph{Deep model-based architectures} (DMBAs),
	such as those based on deep unfolding~\cite{hershey2014deep, monga2021algorithm},
	have recently extended traditional DL to neural network architectures
	that combine the CT forward models and CNN regularizers%
	~\cite{Hauptmann.etal2018, Adler.etal2018, liu2021sgd,Mukherjee.etal2021, liu2022online, wu2017iterative, chun2020momentum, zhou2021limited, huang2019data}.

	\begin{figure*}[t]
		\includegraphics[width=\textwidth]{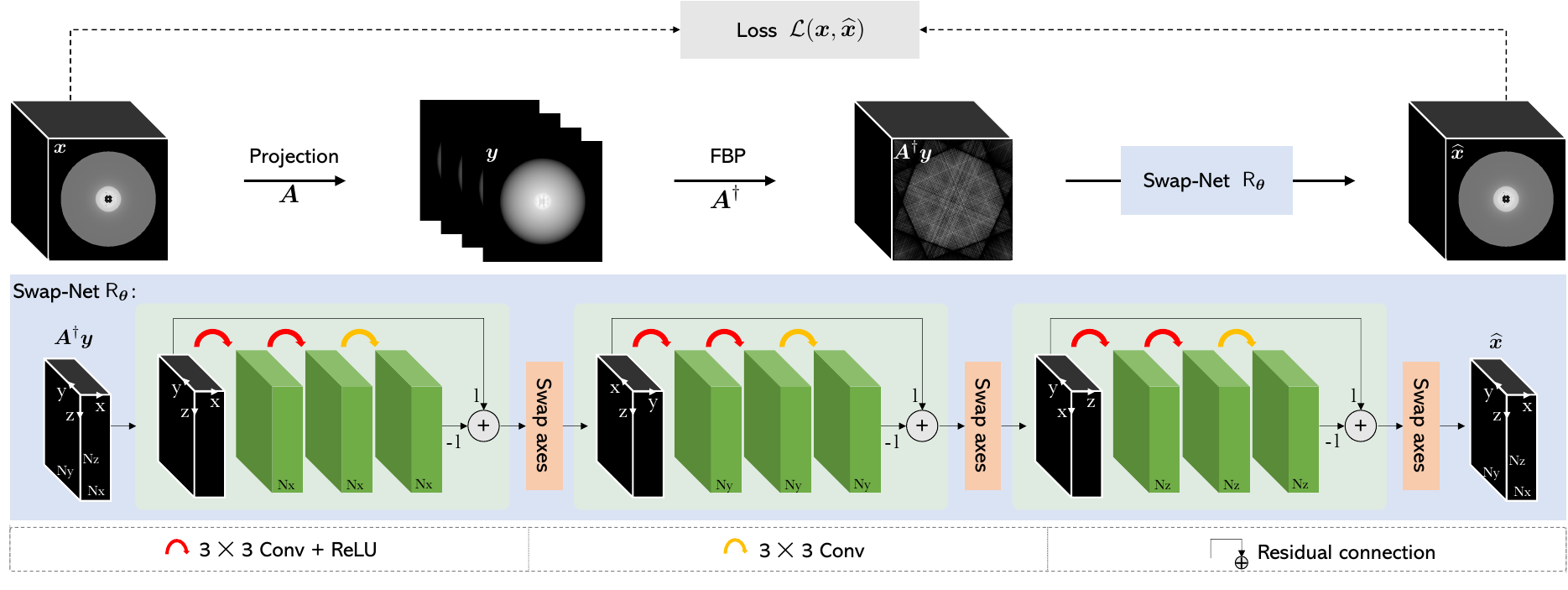}
		\caption{Overview of the proposed Swap-Net framework
			for training an end-to-end deep mapping for 3D CBCT image reconstruction using ICF synthetic radiographs.
			The Swap-Net model $\Rsf_\thetabm$ is implemented as a customized architecture
			mapping the output of FBP to the desired ground-truth 3D images.
			The novel axes-swapping operation in Swap-Net
			allows it to efficiently conduct convolution across all dimensions.
			The whole network is trained end-to-end in a supervised fashion. }
		\label{fig:pipeline}
	\end{figure*}
	Despite the rich literature on DL-based methodologies,
	direct end-to-end 3D CT reconstruction remains a challenging problem
	due to its high memory and computation cost.
	Current schemes typically use a 2D approach,
	where the 3D volume is divided into a series of 2D slices along one or more axes.
	Then each 2D slice is treated as an independent image,
	and a 2D neural network is applied to process each slice individually.
	After processing all 2D slices,
	the outputs are combined to reconstruct the full 3D volume.
	Using a 2D network for 3D reconstruction offers several advantages,
	including computational efficiency, ease of implementation,
	and compatibility with existing 2D CNN architectures and frameworks.
	However, it also suffers from drawbacks
	such as the potential loss of consistency across slices
	and suboptimal performance in capturing complex 3D structures
	compared to dedicated 3D reconstruction approaches~\cite{Lee_2023_ICCV,
		8850991}.
	This paper addresses these issues by presenting a new network---called Swap-Net---%
	for recovering high-quality 3D images from extreme sparse-view measurements.
	Distinct from the fully 3D volume-based approaches and 2D slice-based approaches,
	Swap-Net is developed as a 2.5D CNN
	where 2D convolution operations
	are used to extract correlations across all three dimensions of a 3D volume.
	The key contributions of our work are summarized as follows: 
	\begin{itemize}[leftmargin=*]
		\item
		We present a memory-efficient 2.5D network called Swap-Net
		to handle end-to-end 3D image reconstruction.
		The key component in Swap-Net is the new axes-swapping operation
		that helps combine information along all axes similar to 3D convolution.
		
		\item
		We investigated challenging sparse-view 3D CBCT image reconstruction problems
		with as few as 4 projection views.
		Moreover, we accounted for non-ideal physics
		including blur, scatter, and non-white noise.
		%
		Simulation results demonstrate that the method can restore high-quality 3D volumes across all dimensions,
		outperforming baseline methods both quantitatively and qualitatively
		in terms of artifact-reduction and detail-preservation.
		
		\item
		We conducted additional investigations using Swap-Net, e.g.,
		studying the benefits of the axis-swapping and the impact of the swapping order.
		Our results show that the properly chosen axis-swapping order
		can effectively boost the performance of the network.

		
		
	\end{itemize}
	
	The rest of this paper is organized as follows.
	Section~\ref{sec:background}
	introduces the background and mathematical formulation of the CBCT imaging problem
	and discusses related work.
	Section~\ref{sec:methods}
	presents our proposed approach in detail.
	Section~\ref{sec:experiments}
	explains our experimental setup, presents the results of our comparisons to other algorithms,
	and elaborates upon the analysis of the observations.
	Finally, Section~\ref{sec:conclusion}
	summarizes our work and discusses potential future directions.

	\begin{figure*}[t]
		\includegraphics[width=\textwidth]{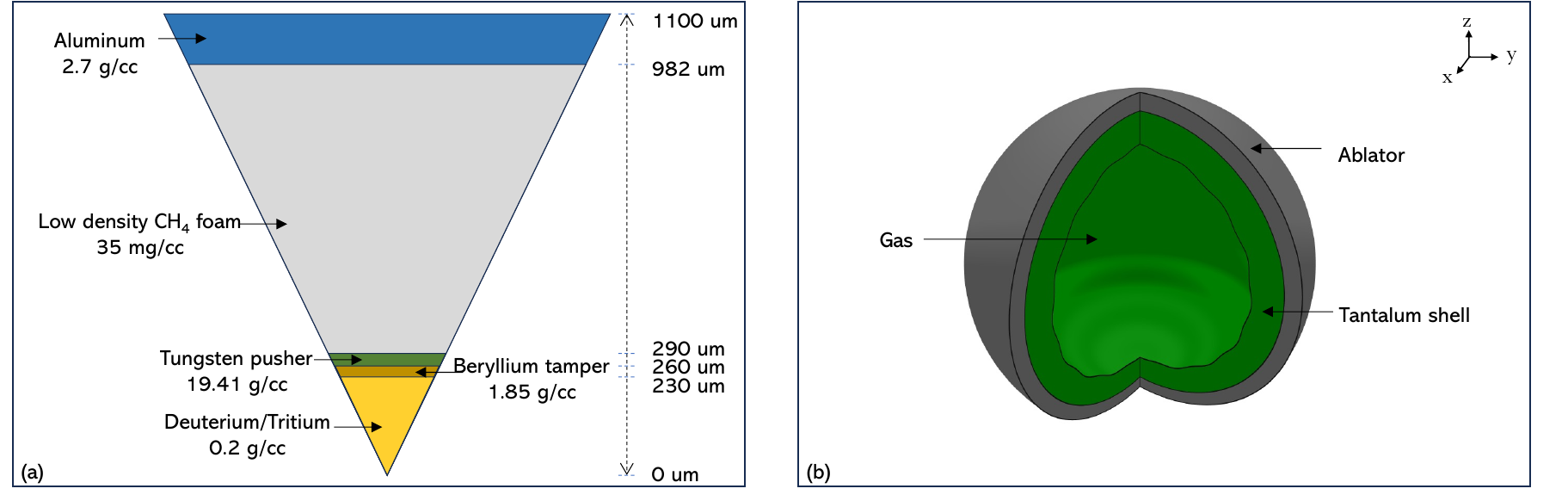}
		\caption{The ICF models: (a) A typical double-shell ICF capsule containing Deuterium/Tritium, Beryllium tamper, Tungsten pusher, low density $\text{CH}_4$ foam, and Aluminum.
			(b) A simplified representation of a ICF implosion capsule containing Deuterium/Tritium \todo{(Ablator)}, Tungsten pusher \todo{(Tantalum)},
			and low density $\text{CH}_4$ foam \todo{(Gas)}~\cite{merritt2019experimental}.}
		\label{fig:double-shell_and_icf}
	\end{figure*}
	\section{Background}
	\label{sec:background}
	
	\subsection{CT Inverse Problem Formulation}
	
	In CT imaging, the relationship between the unknown object $\xbm \in \R^n$
	and the (log) projection
	measurements
	$\ybm\in\R^m$ is commonly expressed as a linear imaging system
	\begin{equation}
		\label{equ:imaging_awgn}
		\ybm = \Abm\xbm + \ebm,
	\end{equation}
	where $\Abm\in\R^{m\times n}$ denotes the measurement operator
	(also known as the forward model or physical model)
	and $\ebm\in\R^m$ denotes the measurement noise
	that is sometimes statistically modeled as additive white Gaussian noise (AWGN).
	The AWGN formulation is a widely used approximation
	for various imaging systems including CT, magnetic resonance imaging (MRI),
	etc.~\cite{thibault2007three, 5484183}.
	
	Scatter is another practical corruption that arises in CT imaging
	due to interactions between X-ray photons and objects.
	When X-ray photons encounter the object,
	some of them undergo scattering rather than being absorbed or passing straight through.
	Since scattered photons have undergone direction changes,
	they do not provide accurate information about the original object attenuation along the X-ray path.
	Scattered photons can reach the detector
	and contribute to errors that reduce the quality of the reconstructed image.
	Often the post-log scatter-corrupted CBCT projection measurements $\ybm$ are modeled as 
	\begin{equation}
		\label{equ:imaging_scatter}
		\ybm = -\log \left(\frac{\Phi(\Ibm_0e^{-\Abm\xbm})}{\Ibm_0}\right)\ , 
	\end{equation}
	where $\Ibm_0$ denotes the reference intensity from the source,
	$\Phi$ is a nonlinear function that models
	the non-ideal physics including blur and scatter corruption
	(see Section~\ref{sec:experiments}-B for details),
	and $\log$ is applied pixelwise.
	Choices for modeling the scatter component of the function $\Phi$
	in the literature
	include kernel convolution with the direct signal
	followed by Poisson noise~\cite{sun_improved_2010, mccann_local_2021}.
	For any noise model,
	the goal is to reconstruct the image volume \x
	from the projection data \y.

	\subsection{Related Work for CT Reconstruction}
	\label{sec:dl-reconstruction}
	
	As a 3D-imaging technique,
	CBCT imaging offers many benefits in clinical, industry, and research.
	However, due to factors such as computation cost, scatter, noise, limited measurements,
	and discrepancies in the forward operator model,
	significant challenges emerge
	when attempting to efficiently and accurately reconstruct the 3D CT images
	outlined in~\eqref{equ:imaging_awgn} and~\eqref{equ:imaging_scatter}%
	~\cite{delaney1998globally, yu2002edge, pasha2023computational}.
	Classical approaches tackle CT reconstruction
	by formulating it as a regularized optimization problem
	\begin{equation}
		\label{equ:optimization}
		\xbmhat = \argmin_{\xbm\in\R^n} \{\df(\xbm) + \reg(\xbm)\} \ ,
	\end{equation}
	where $\df$ is the data-fidelity term
	that quantifies the consistency with the measured data $\ybm$,
	and $\reg$ is a regularizer that enforces a prior knowledge on the unknown image $\xbm$.
	For example, two widely-used data-fidelity and regularization terms in imaging
	are the least-squares and total variation (TV) terms:
	\begin{equation}
		\label{equ:optimization_tv}
		\df(\xbm) = \frac{1}{2}\norm{\ybm-\Abm\xbm}^2_2\quad \mathrm{and}\quad \reg(\xbm)=\tau\norm{\Dbm\xbm}_1\ ,
	\end{equation}
	where $\tau > 0$ controls the regularization strength
	and $\Dbm$ denotes the discrete gradient operator~\cite{Rudin.etal1992}.
	Many handcrafted regularizers similar to TV
	have also been applied to sparse-view CT reconstruction problems%
	~\cite{sidky2006accurate, zhang2016low,herman2008image,chen2008prior,bian2010evaluation,ramani2011splitting}.
	Beyond handcrafted priors,
	recent work has also explored the use of learned priors, e.g.,%
	~\cite{pfister2014model, zhang2016low, zheng2018pwls, liu2021sgd, liu2022online}.
	
	DL has gained great popularity for solving CT inverse problems
	due to its excellent performance~\cite{dong2019deep, kim2019extreme, montoya2022reconstruction}.
	A widely used supervised DL approach is based on training a CNN
	to map a corrupted image to its clean target
	\cite{jin:17:dcn}.
	For example, prior work on DL for CBCT trains a CNN
	to map FBP reconstructed images to the corresponding ground-truth images.
	In particular, for CBCT where the target images are 3D,
	due to the memory limits,
	the network training is typically done in a slice-by-slice manner,
	where the 3D volumes are sliced into 2D images along a certain axes
	and the loss is optimized on the given slices%
	~\cite{han2018deep, guan2020limited,liu2021sgd, liu2022online}.
	However, due to the lack of global information,
	a 2D slice-based approach cannot capture complex 3D structures
	as well as dedicated 3D reconstruction approaches~\cite{xu2022learning}.
	An alternative method is to divide the whole volume into small 3D patches,
	feed the patches to the 3D network,
	and then fuse the reconstructed patches together
	(see reviews in~\cite{karimi2016patch}).
	While such 3D patch-based approach can extract and establish features in all dimensions within patches, it cannot model global correlations and the fusion of patches in forming the whole volume
	requires additional attention to boundary artifacts~\cite{majee20194d}.
	
	\begin{figure*}[t]
		\begin{center}
			\includegraphics[width=.5\textwidth]{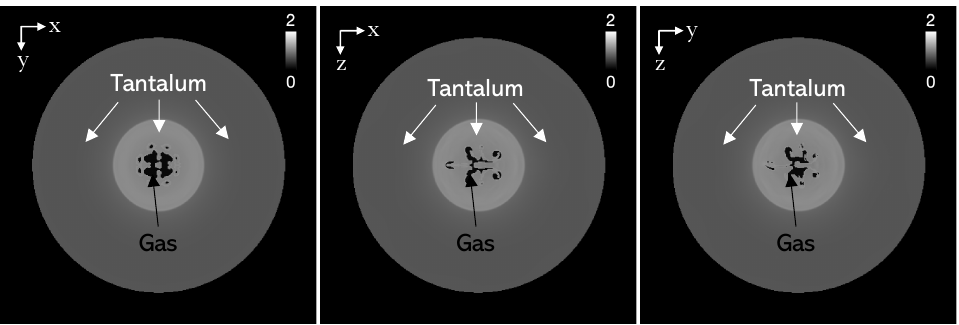}
		\end{center}
		\caption{Central slices along each dimension
			of an exemplar 3D ICF object generated for an ICF double shell simulation in our dataset.
			The two materials that form the object, namely gas and metal,
			are labeled in each image.
			The images presented here were normalized by the mass attenuation factor to the range of [0, 2]
			for good visualization (same in the rest of the paper). }
		\label{fig:samples}
	\end{figure*}

	\subsection{Our Contribution}
	\label{sec:3d cbct}
	
	This work contributes to the memory-expensive area
	of efficient 3D CBCT reconstruction using DL methods.
	We introduce a memory-efficient 2.5D network, called Swap-Net,
	that refines
	3D images
	reconstructed
	from artifact-corrupted radiographs.
	Swap-Net addresses in an end-to-end fashion several common sources of image artifacts,
	including those due to sparse view sampling, measurement noise, and photon scattering.
	We extensively test the performance of Swap-Net,
	validating that it can be used as an effective end-to-end mapping tool for 3D CBCT image reconstruction.

	\section{Proposed Method}
	\label{sec:methods}
	
	We propose Swap-Net as an end-to-end mapping network
	that can handle 3D inverse problems like CBCT reconstruction.
	Fig.~\ref{fig:pipeline} shows the training pipeline (top) and architecture (bottom) of Swap-Net.
	As illustrated in the top part of Fig.~\ref{fig:pipeline},
	given the corrupted CBCT projections $\ybm$,
	Swap-Net $\Rsf_\thetabm$ takes the FBP reconstructed images $\Abm^\dagger\ybm$ as its input,
	and maps the whole volume to the desired 3D output
	$\xbmhat \defn \Rsf_\thetabm(\Abm^\dagger\ybm)$.
	Here, $\thetabm$ represents the parameters of Swap-Net,
	and $\Abm^\dagger$ denotes the FBP reconstruction operation.
	Swap-Net training seeks to minimize the loss $\mathcal{L}$
	between $\xbmhat$ and the ground truth $\xbm$ over a training set
	consisting of $J$ samples
	to obtain the optimized parameters $\thetabm^\ast$
	\begin{equation}
		\thetabm^\ast = \argmin_{\thetabm} \sum_{j = 1}^J \mathcal{L}(\xbm_j, \Rsf_\thetabm(\Abm^\dagger\ybm_j)\ ,
		\label{eq:loss_train}
	\end{equation}
	where $\mathcal{L}$ denotes the loss function
	that measures the discrepancy between the predictions of the Swap-Net and the ground truth.
	
	\begin{table*}[t]
		\centering
		\footnotesize
		\begin{threeparttable}
			\caption{Quantitative evaluation of Swap-Net and baseline methods
				averaged on the test dataset for different numbers of projection views.
				Swap-Net contains the fewest network parameters (PN) and uses moderate amount of GPU running  memory (RM) yet achieved the highest SNR and SSIM compared with all the baseline methods
				across different projection settings.}
			\label{tab:evaluation}
			\renewcommand\arraystretch{1}
			\setlength{\tabcolsep}{3.8pt}
			\begin{tabular}{lccccccccccccc} 
				\toprule
				\textit{Settings} & & \multicolumn{6}{c}{AWGN} & \multicolumn{6}{c}{Scatter} \\ 
				\cmidrule(rl){3-14}
				
				\textit{Metric} & PN (Million) / RM (GB) & \multicolumn{3}{c}{SNR (dB)} & \multicolumn{3}{c}{SSIM}
				& \multicolumn{3}{c}{SNR (dB)} & \multicolumn{3}{c}{SSIM} \\
				\cmidrule(rl){2-2}\cmidrule(rl){3-8}\cmidrule(rl){9-14} 
				
				\textit{Views} & --- & 4 & 8 & 16 & 4 & 8 & 16 & 4 & 8 & 16 & 4 & 8 & 16 \\
				\cmidrule(rl){2-2}\cmidrule(rl){3-5}\cmidrule(rl){6-8}\cmidrule(rl){9-11}\cmidrule(rl){12-14}
				FBP &---& 9.04 & 13.19 & 15.53 & 0.63 & 0.68 & 0.72 & 9.44 & 11.57 & 12.03 & 0.64 & 0.7 & 0.75 \\
				TV &---& 14.01 & 17.05 & 18.48 & 0.57 & 0.6 & 0.82 & 9.79 & 11.49 & 12.08 & 0.73 & 0.66 & 0.76 \\
				2D U-Net &\; 50.26 / \textbf{[0.77, 1.66]} & 20.77 &20.93 & 20.94 &  0.95 & 0.95 & 0.95 & 17.01 & 17.77 & 20.30 & 0.89 & 0.91 & 0.95  \\
				3D U-Net & 150.75 / [2.27, 9.25] & 26.59 & 27.32 & 27.62 & 0.99 & 0.99 & 0.99 & 19.18 & 19.28 & 20.42 & 0.99 & 0.99 & 0.99 \\
				Swap-Net (Ours) &\; \textbf{16.26} / [1.25, 5.91] & \textbf{28.22} & \textbf{28.58} & \textbf{28.83} & \textbf{0.99} & \textbf{0.99} & \textbf{0.99} & \textbf{25.41} & \textbf{25.46} & \textbf{25.60} & \textbf{0.99} & \textbf{0.99} & \textbf{0.99} \\
				\bottomrule
			\end{tabular}
		\end{threeparttable}
	\end{table*}
	
	\begin{figure*}[t]
		\includegraphics[width=\textwidth]{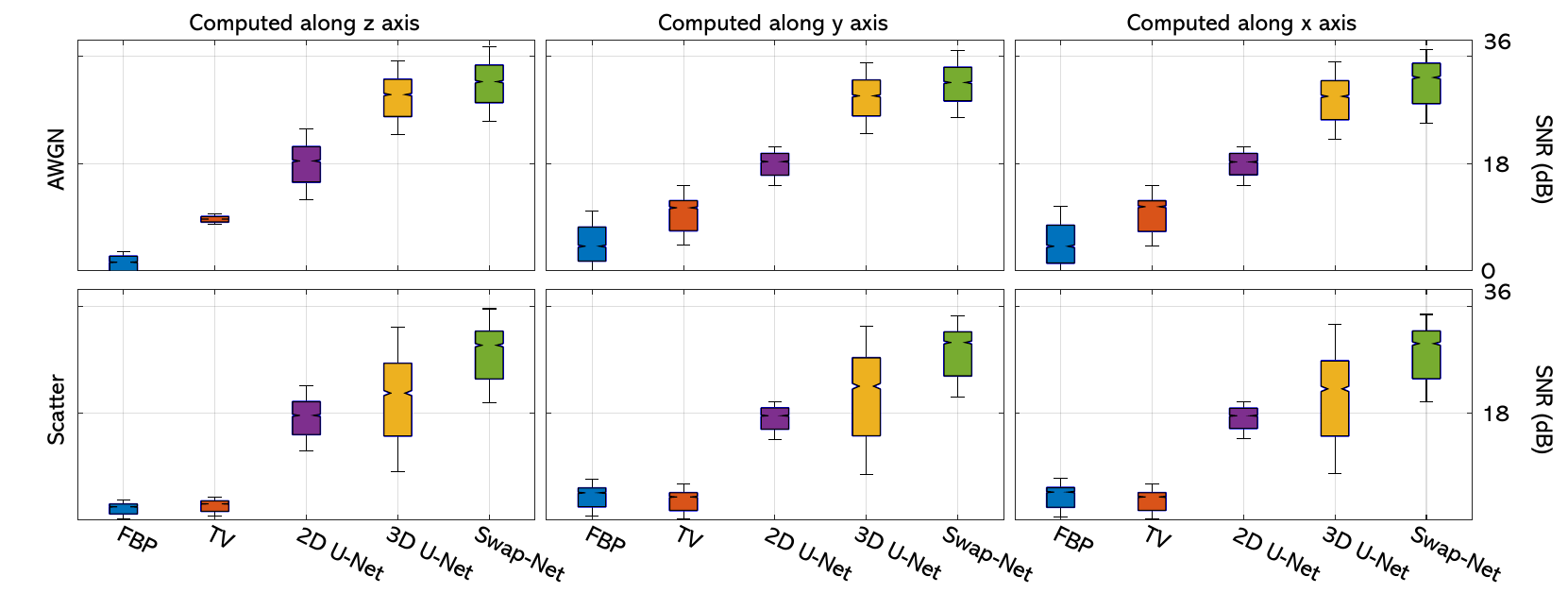}
		\caption{Statistical summary of SNR values
			for different reconstruction methods
			evaluated on 2D slices along each dimension
			taken from our test set.
			Plots in the first and second row
			correspond to the the results with 4 projection views under AWGN
			and non-ideal physics including blur and scatter and non-white noise corruptions, respectively.}
		\label{fig:statistic}
	\end{figure*}
	
	\begin{figure*}[t]
		\includegraphics[width=\textwidth]{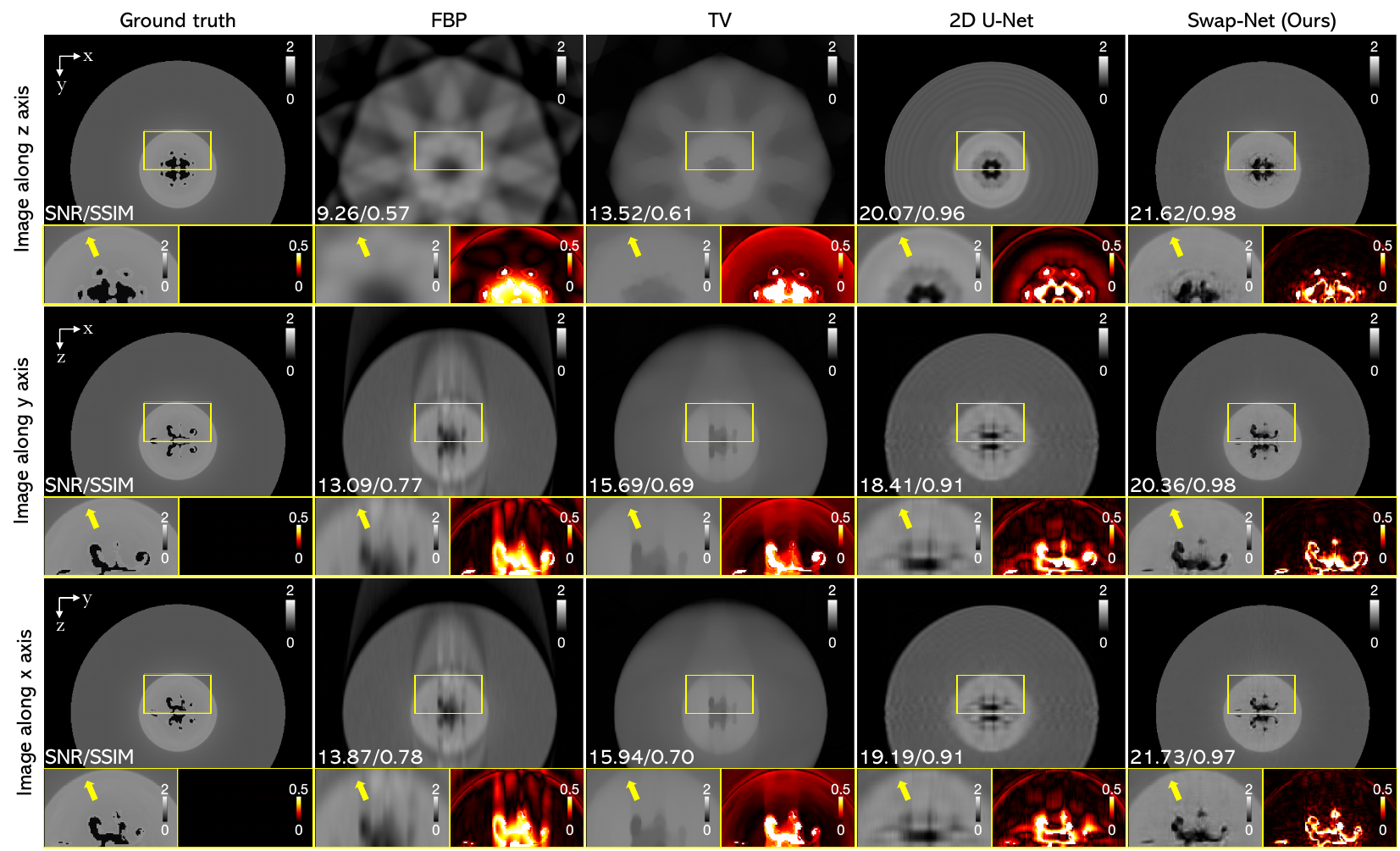}
		\caption{Visual evaluation of Swap-Net and baseline methods
			on an exemplar ICF double shell test simulation with $4$ projection views under AWGN corruption.
			Each row shows the middle slice of the whole 3D object along $z$, $y$ and $x$ axes, respectively.
			The bottom part of each image provides the SNR and SSIM values
			and representative $2\times$ zoomed-in regions and their error maps with respect to the ground truth.
			Arrows in the zoomed-in plots highlight sharp edges that are well reconstructed using Swap-Net.
			Note the excellent quantitative and qualitative performance of Swap-Net
			for both artifact correction and detail preservation.}
		\label{fig:compare_methods_awgn}
	\end{figure*}

	When the input images $\Abm^\dagger\ybm$ are of size $N_x \times N_y \times N_z$,
	Swap-Net works in a 3D-to-3D manner
	to produce a whole volume estimate $\xbmhat$
	having the same dimension as its input
	without slicing and assembling the volume.
	The efficiency of Swap-Net in facilitating 3D image reconstruction
	hinges upon the novel and efficient design of its architecture.
	As illustrated in the bottom part of Fig.~\ref{fig:pipeline},
	Swap-Net is a cascade of three repeating blocks,
	each consisting of two convolutional layers (Conv)
	followed by Rectified Linear Unit (ReLU) activations,
	one additional convolutional layer, and a residual connection.
	The convolutional kernels across all layers are uniformly set to a size of $3 \times 3$ with a stride of 1. The channel dimensions of the hidden convolutional layers are set to $N_x$, $N_y$, and $N_z$ for the first, second, and final blocks, respectively, corresponding to the dimensions of the 3D input volume along the $x$, $y$, and $z$ axes.
	A distinctive aspect of Swap-Net lies in its use
	of the \emph{axes-swapping} operation after each block.
	This operation sequentially reorients the channel dimension to the $x$, $y$, and $z$ axes,
	facilitating focused 2D convolutions across the $yz$, $xz$, and $xy$ planes, respectively.
	This strategic approach enables the network to perform artifact reduction axis by axis,
	thereby ultimately yielding a high-fidelity 3D reconstruction
	that maintains consistency across all dimensions. 
	
	The key novelty of our method is that,
	to the best of our knowledge,
	this is the first work presenting such a memory-efficient 2.5D cascade network
	based on axes-swapping operations.
	Different from the traditional slice-by-slice mapping methods,
	Swap-Net instead relies on the axis-by-axis reconstruction,
	which can be particularly useful when the 3D volume is not uniformly corrupted along each dimension.
	For example, in CBCT imaging, since the projections are produced
	by rotating the X-ray beam along a certain axis,
	e.g., the $z$ axis in our experiments,
	the insufficient attenuation information along $z$
	usually leads to lower-quality images in the $xy$ plane.
	The cascading axes-swapping operations in Swap-Net
	allow it to effectively process the reconstruction along axis $z$
	after accumulating more information along axis $x$ and axis $y$,
	therefore enforcing the global consistency of the reconstructed 3D volume. Moreover, although the output of Swap-Net is the whole 3D volume,
	it does not involve any computationally expensive 3D convolutions.
	Instead, it is simply based on 2D convolutions
	where the convolution is looped over all axes of a 3D volume.
	Thus, Swap-Net
	overcomes the suboptimal performance of slice-based 2D CNNs
	that disregard the information across slices.
	On the other hand, it also bypasses the expensive computation cost of 3D networks,
	facilitating solving practical 3D imaging problems.

	\begin{figure}[t]
		\begin{center}
			\includegraphics[width=.5\textwidth]{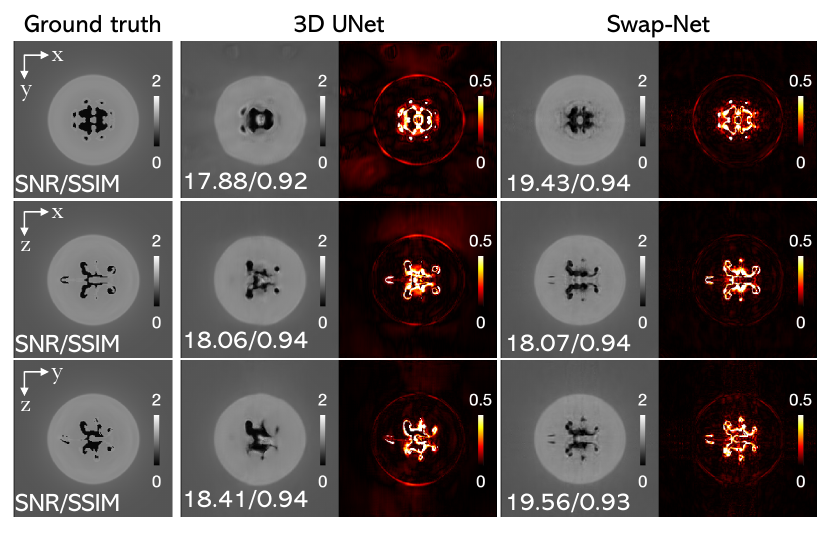}
		\end{center}
		\caption{Visual evaluation of 2.5D Swap-Net and 3D U-Net
			on an exemplar ICF double shell test simulation with $4$ projection views under AWGN corruption.
			Each row shows the middle slice of the central region of the 3D object
			and the corresponding error maps with respect to the ground truth along $z$, $y$ and $x$ axes, respectively.
			The bottom part of each image provides the SNR and SSIM values.
			With only about $1/10$ of the parameters of 3D U-Net,
			Swap-Net still achieves better quantitative and qualitative performance.}
		\label{fig:compare_3d}
	\end{figure}[t]
	
	\begin{figure*}[t]
		\includegraphics[width=\textwidth]{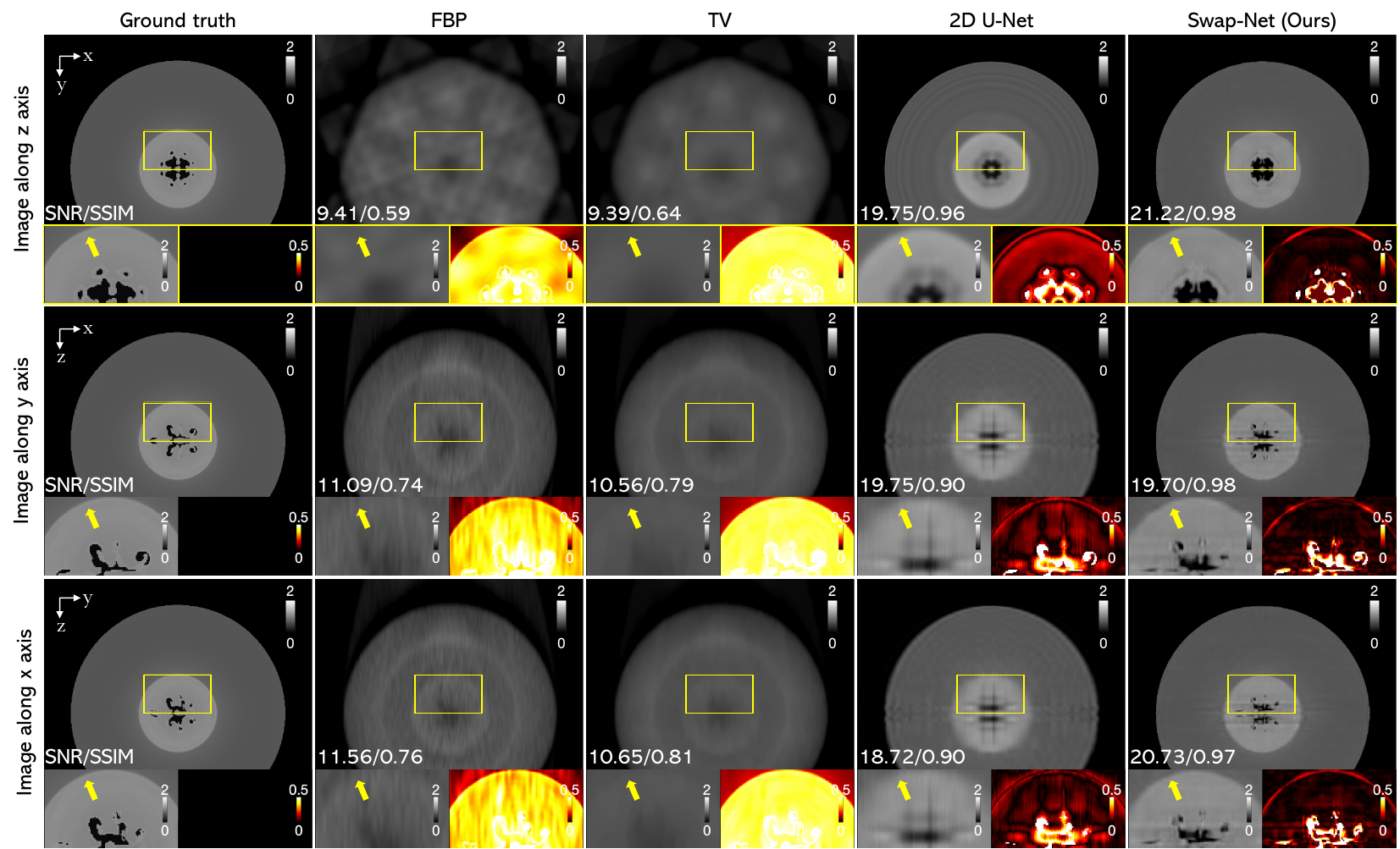}
		\caption{Visual evaluation of Swap-Net and baseline methods
			on an exemplar ICF double shell test simulation 
			with $4$ projection views under scatter corruption.
			Each row shows the middle slice of the whole 3D object along $z$, $y$ and $x$ axes, respectively.
			The bottom part of each image provides the SNR and SSIM values, and representative $2\times$ zoomed-in regions and their error maps with respect to the ground truth.
			Arrows in the zoomed-in plots highlight sharp edges that are well reconstructed using Swap-Net.
			Note the excellent quantitative and qualitative performance of Swap-Net
			for both artifacts correction and detail preservation.}
		\label{fig:compare_methods_scatter}
	\end{figure*}

	\section{Experimental Validation}
	\label{sec:experiments}
	
	This section presents numerical results
	that demonstrate the ability of Swap-Net to provide high-quality 3D reconstructions
	from sparse-view 2D projections of ICF double-shell capsules
	as depicted in a representative double shell shown in Fig.~\ref{fig:double-shell_and_icf}.
	In particular, we examine Swap-Net under two different practical noise conditions,
	including AWGN corruption and nonlinear photon scattering corruption,
	to show its ability to solve challenging CBCT imaging problems.

	\subsection{Preparation of 3D Dataset}
	
	The emergence of Inertial Confinement Fusion (ICF) as a potential power source
	has been a major impetus for the continued examination of ICF implosion dynamics.
	One promising ICF configuration is a double-shell capsule,
	shown in Fig.~\ref{fig:double-shell_and_icf} (a),
	that employs a high Z metallic shell that is imploded onto a gas-filled cavity
	via radiation to achieve fusion conditions.
	Both manufacturing as well as drive asymmetries may lead to hydrodynamic instabilities
	that can degrade ICF performance.
	Consequently, quantifying and understanding these instabilities
	is crucial to the continued success of ICF.
	To this end, radiography plays an essential role
	in elucidating the behavior of the metallic shell
	and quantifying the impact of the asymmetries on ICF performance.
	
	To further simplify the problem,
	we examine the implosion of a single shell made of tantalum,
	as this configuration enables the salient features to be captured in the density field,
	i.e., a complex gas metal interface without needing to increase the simulation complexity. 
	As such we train and test our method with ICF capsules shown
	in Fig.~\ref{fig:double-shell_and_icf} (b)
	to examine shock propagation and instability growth created
	using prescribed perturbations on the shell interior surface.
	All simulations were performed using computational fluid dynamics software. 
	

	In particular, our ICF capsules simulations were performed 
	on a $440\times 440\times 440$
	uniform Cartesian grid
	over $[0, L] \times [0, L]\times [0, L]$,
	where 
	$L = 341$~$\mu$m.
	The uniform grid cell size is
	$\Delta x = \Delta y = \Delta z = \frac{440}{L}$.
	We used
	108 3D objects with different parameters,
	e.g., initial 3D perturbations, material properties, and/or temporal slice
	where each case represents a distinct dynamic hydrodynamic configuration.
	Fig.~\ref{fig:samples} shows an exemplar object from our datasets
	with gas and tantalum labeled.
	In particular, the mass attenuation coefficient of gas is
	$\xi_{(\rm gas)}=9.40\;\text{cm}^2/\text{g}$,
	and tantalum is $\xi_{(\rm tantalum)}=13.03\;\text{cm}^2/\text{g}$, in the energy range of interest here.
	Each object has dimensions $448 \times 448 \times 448$
	with voxel size $250\times 250 \times 250$ um$^3$.
	These 108 objects were split into 90, 18, and 18 for training, validation, and testing, respectively.

	\subsection{Generation of Radiographs}
	\label{subsec:Generation of Radiographs}
	
	The direct radiographic signals from the area mass were simulated
	by rotating the X-ray source along axis $z$ with base intensity $I_0=3.201 \times 10^{-4}$.
	We tested the performance of Swap-Net on sparse-view CBCT reconstruction with 4, 8 and 16 views.
	The dimensions of the 2D projections were $200 \times 200$ with resolution $2000\times 2000\, \mu\mathrm{m}^2$.
	These CBCT views were generated using the ODL package~\cite{jonas_adler_2017_249479},
	and all views were equally spaced over 180 degrees.  
	
	We generated the radiographs under two different corruption scenarios,
	namely AWGN,
	as modeled in~\eqref{equ:imaging_awgn},
	and non-ideal physics including blur
	and scatter and non-white noise corruption,
	as modeled in~\eqref{equ:imaging_scatter}, respectively.
	For AWGN, the simulated corrupted radiographs included the addition of random AWGN
	corresponding to an input SNR of 40dB to the clean ones.
	For our non-idea physics investigation,
	we modeled the total transmission or the noisy radiograph function $\Phi$
	as the sum of the blurred radiograph, scatter, and noise as follows:
	\begin{equation}
		\Phi :=\Dblur + \Dscat + \Bscat + \bm{\eta}.
		\label{eq:noisy_radiograph}
	\end{equation}
	Let \D denote the uncollided radiation
	incident on the detector plane.
	The blurred direct radiation component is given by
	\begin{equation}
		\Dblur = \D \ast \Gblur(\sblur) \ast \bm{\phi}_\text{db}.
		\label{eq:detector_source_blur}
	\end{equation}
	The source blur \Gblur is given by a 2D Gaussian kernel
	with deviation \sblur chosen randomly between 1 and 3 pixels
	with an accompanying random orientation between 5 and 26 degrees.
	This signal was then convolved with a detector blur using another kernel $\bm{\phi}_\text{db}$.

	To address the scatter radiation, we included two  scatter components.
	The first was a correlated scatter component given by 
	\begin{equation}
		\Dscat = \kappa \D \ast \Gscat(\sscat).
		\label{eq:scatter}
	\end{equation}
	Here we convolved the direct radiograph
	with a 2D Gaussian filter scatter kernel \Gscat
	having standard deviation \sscat between 10 and 30 pixels for the kernel,
	with a scaling factor $\kappa$ between \todo{0.1 and 0.3}.
	We also added a background scatter field \Bscat,
	which is another essential component of scatter affecting radiographic measurements.
	Physically, this term represents scatter from our object that is reflected
	by nearby surrounding objects, e.g., ground and walls,
	which are particularly difficult to model.
	This field was modeled with a polynomial of order $n$ given as
	\begin{equation}
		\Bscat(x,y) = \sum_{i=0}^n a_i x^i + b_iy^i, 
		\label{eq:background_scatter}
	\end{equation}
	where $x$ and $y$ denote spatial coordinates
	and $a_i$ and $b_i$ denote the coefficients of the polynomial.
	We chose
	the coefficients of the background scatter field
	such that the level was randomly between 0.5 and 1.5 times
	the mean signal level in the center of the image
	and the tilt was between -10\% and 10\%.

	We modeled
	gamma and photon noise as Poisson noise denoted by
	$\bm{\eta}_g^\text{Po}$ and $\bm{\eta}_p^\text{Po}$, respectively.
	The means of the two distributions were proportional to the total signal
	$\Dblur + \Dscat + \Bscat$ (with a scaling for each noise).
	The noise components were convolved with respective kernels
	$\bm{\phi}_g$ and $\bm{\phi}_p$ to give the total (colored) noise $\bm{\eta}$ as follows:
	\begin{equation}
		\bm{\eta}=\kappa_g(\bm{\eta}^\text{Po}_g \ast \bm{\phi}_g) + \kappa_p (\bm{\eta}^\text{Po}_p \ast \bm{\phi}_p),
		\label{eq:poisson_noise}
	\end{equation}
	where $\kappa_g$ and $\kappa_p$ are scaling coefficients
	for the gamma and photon noise components, respectively.
	The level of the gamma noise was randomly set in the range of (39,000, 50,000)
	and the level of the photon noise was randomly set in the range (350, 450).
	All random parameters were generated independently for each radiograph,
	so each radiograph was corrupted with different random noise and scatter realizations.

	\subsection{Baseline Methods and Training Settings}
	
	We considered several well-known algorithms as baseline methods for CBCT image reconstruction,
	including \emph{FBP}, \emph{TV}~\cite{Rudin.etal1992},
	\emph{2D U-Net}~\cite{Ronneberger.etal2015}, and \emph{3D U-Net}~\cite{cciccek20163d}.
	FBP and TV are traditional methods that do not require training,
	while other methods are all DL methods with publicly available implementations.
	The FBP method was performed with the Hann filter,
	and the relative cutoff frequency for the filter was set to $0.3$.
	We used \texttt{fminbound} in the \texttt{scipy.optimize} toolbox
	to identify the optimal regularization parameter $\tau$ for TV at the inference time.
	We trained all DL methods on the FBP reconstructed images to handle CBCT reconstruction.
	For 3D U-Net, we trained the model with 3D patches with patch size set to $112 \times 112 \times 112$,
	and fused patches
	to form its final reconstruction.
	We used the $\ell_2$ loss function for all training approaches,
	and set the learning rate to $0.001$ and used Adam~\cite{Kingma.Ba2015} as our training optimizer.
	All models were trained for 1000 epochs, at which point stable convergence was observed.
	We evaluated reconstruction performance
	using two widely-adopted metrics:
	signal to noise ration (SNR) in dB and structural similarity index measure (SSIM)
	from \texttt{skimage.metrics} toolbox.
	Models that achieved the best performance on our validation dataset were selected for inference.

	\begin{figure*}[t]
		\includegraphics[width=\textwidth]{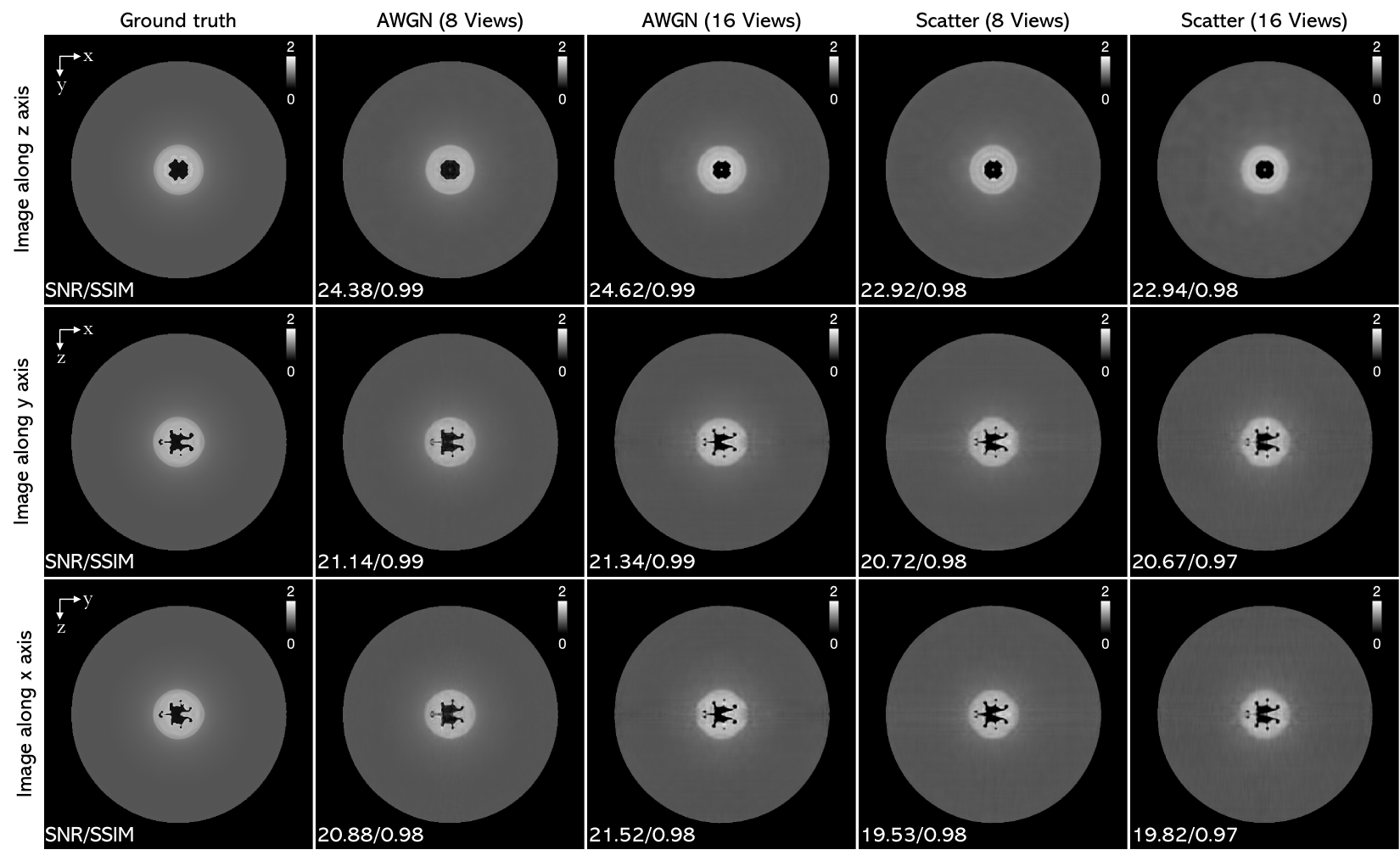}
		\caption{Visual evaluation of Swap-Net on an exemplar ICF double shell test simulation
			with $8$ and $16$ projections views under AWGN and \todo{non-ideal physics including blur
				and scatter and non-white noise (labeled as Scatter)} corruptions.
			Each row shows the middle slice of the whole 3D object along $z$, $y$ and $x$ axes, respectively.
			The bottom-left corner of each image provides the SNR and SSIM values with respect to the ground truth.
			Note the consistently good performance of Swap-Net for different projection views and noise corruptions.}
		\label{fig:compare_views}
	\end{figure*}
	
	\begin{figure*}[t]
		\includegraphics[width=\textwidth]{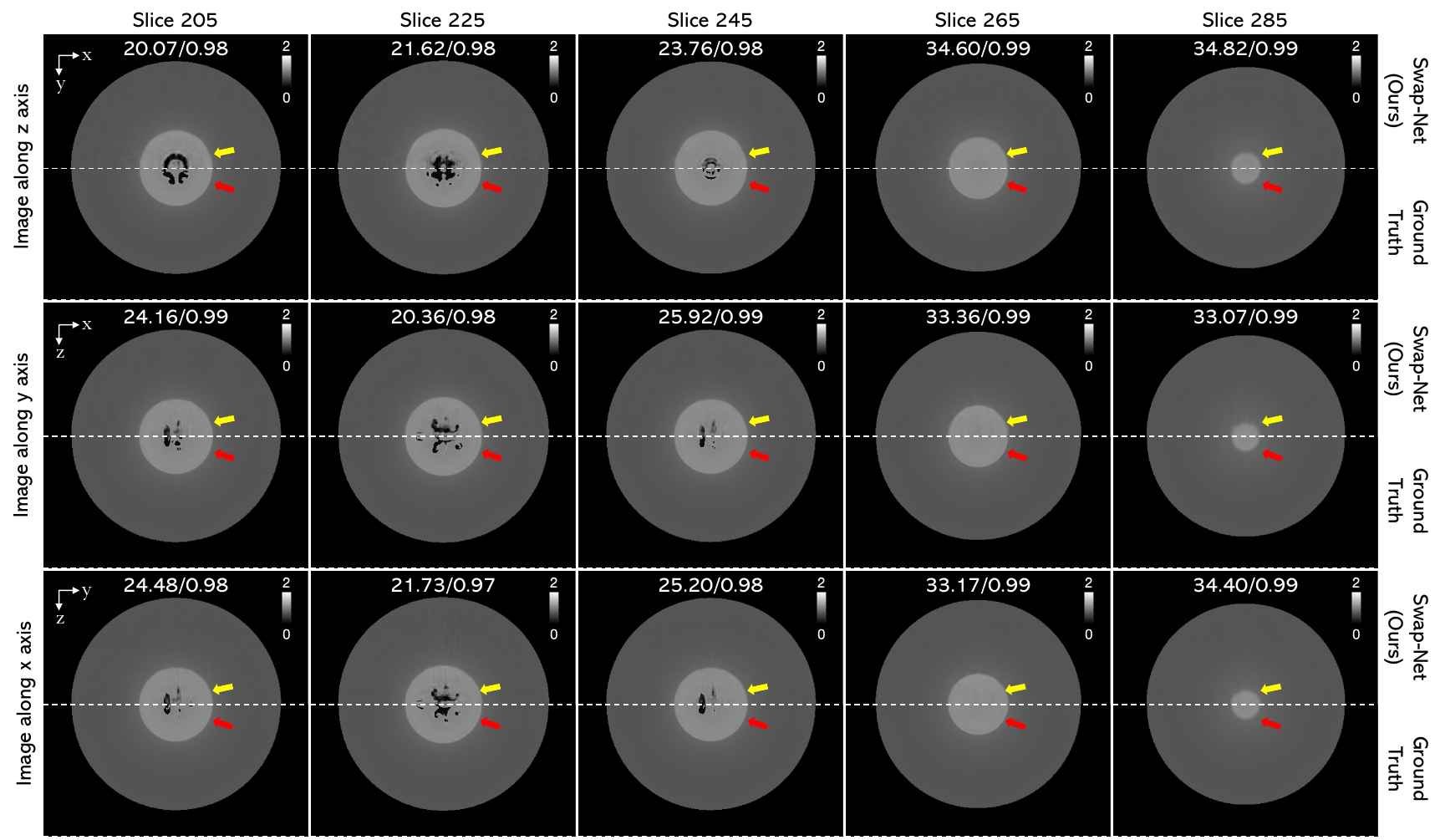}
		\caption{Visual evaluation of Swap-Net across different slices
			on an exemplar ICF double shell test simulation
			with $4$ projection views under AWGN corruption.
			Each row shows different slices of the whole 3D object along $z$, $y$ and $x$ axes, respectively.
			In each row, the images to the top of the dashed line are the  reconstructed images from Swap-Net,
			while the images to the bottom are ground truth.
			The top-middle part of each image provides the SNR and SSIM values with respect to the ground truth.
			Arrows in the plots highlight sharp edge regions that are well reconstructed using Swap-Net.
			Note the consistently good performance of Swap-Net across different slices of a 3D object.}
		\label{fig:compare_slices_awgn}
	\end{figure*}
	
	\subsection{Results and Analysis}
	We first compared the performance of Swap-Net with baseline methods.
	Table~\ref{tab:evaluation}
	summarizes the averaged quantitative evaluation of Swap-Net and baseline methods
	on our testing dataset with different numbers of projection views.
	These numerical results were evaluated on the whole 3D volume
	for both AWGN corruption and \todo{non-ideal
		physics including blur and scatter and photon noise corruption}.
	Swap-Net consistently outperformed the baseline methods,
	leading to the best SNR and SSIM in different scenarios.
	As a reference for model complexity,
	Table~\ref{tab:evaluation} also presents the model size in terms of the number of parameters (PN). Despite obtaining significantly enhanced performance,
	Swap-Net only uses about $1/3$ as many parameters as 2D U-Net
	and $1/10$ as many parameters as 3D U-Net. We also reported the GPU running memory (RM) usage\footnote{The RM usage is presented in the format of [averaged running GPU memory usage, peak running GPU memory usage]. The RM usage for each model were computed by running experiments with batch size of 1 and projection views of 4.} in table~\ref{tab:evaluation}. \todo{Note that Swap-Net processes the entire 3D volume during training, whereas the patch-based 3D U-Net processes  $1/64$ of the volume, and the slice-based 2D U-Net processes $1/448$ of the volume. Despite these differences in data-related RM demands, Swap-Net’s overall RM usage is still lower than that of the 3D U-Net and only slightly higher than the 2D U-Net, making it a memory-efficient solution in practice.} To further evaluate the performance of the reconstruction along each of dimension of the 3D object,
	Fig.~\ref{fig:statistic} summarizes the statistical evaluation for slice-wise reconstruction
	for both AWGN corruption and \todo{non-ideal
		physics including blur and scatter and photon noise} corruptions.
	Swap-Net achieved consistently good reconstruction performance for 2D image slices
	along all three dimensions, thanks to the axes-swapping operation in our network design.

	Fig.~\ref{fig:compare_methods_awgn}
	presents visual comparisons from different methods on an exemplar testing data
	under AWGN corruption with 4 views.
	Swap-Net outperformed the baseline methods
	both in terms of removing artifacts and maintaining sharpness.
	The excellent performance demonstrates that Swap-Net
	can remove disturbing artifacts
	while retaining detailed structural information.
	Such capability is notable for a network having only 9 convolution layers.
	Because the CBCT projections were simulated by rotating along the $z$ axis,
	it is challenging to reconstruct images along $z$
	especially with sparse-view projections
	(e.g., see the comparatively worse FBP reconstruction in $(x,y)$ plane
	in Fig.~\ref{fig:compare_methods_awgn}).
	Swap-Net overcomes such asymmetric artifacts by performing cascading convolutions along all axes,
	resulting in the comparatively consistent reconstruction along all dimensions.
	Fig.~\ref{fig:compare_3d}
	compares Swap-Net and 3D patch-based U-Net methods;
	to avoid the influence of the edge artifacts, only the central region of the reconstructed object is presented.
	In Fig.~\ref{fig:compare_3d},
	Swap-Net performed better than the 3D U-net.
	Fig.~\ref{fig:compare_methods_scatter}
	demonstrates the improved performance of Swap-Net compared with various baseline methods
	under \todo{non-ideal
		physics including blur and scatter and photon noise corruption}.
	While the baseline methods obviously suffer from scatter corruption,
	Swap-Net successfully reduced the artifacts,
	leading to a similar good quantitative and qualitative performance as in the AWGN case.
	Fig.~\ref{fig:compare_views} presents the results of additional investigations
	with the baseline methods using 8 and 16 views.
	
	\begin{figure*}
		\includegraphics[width=\textwidth]{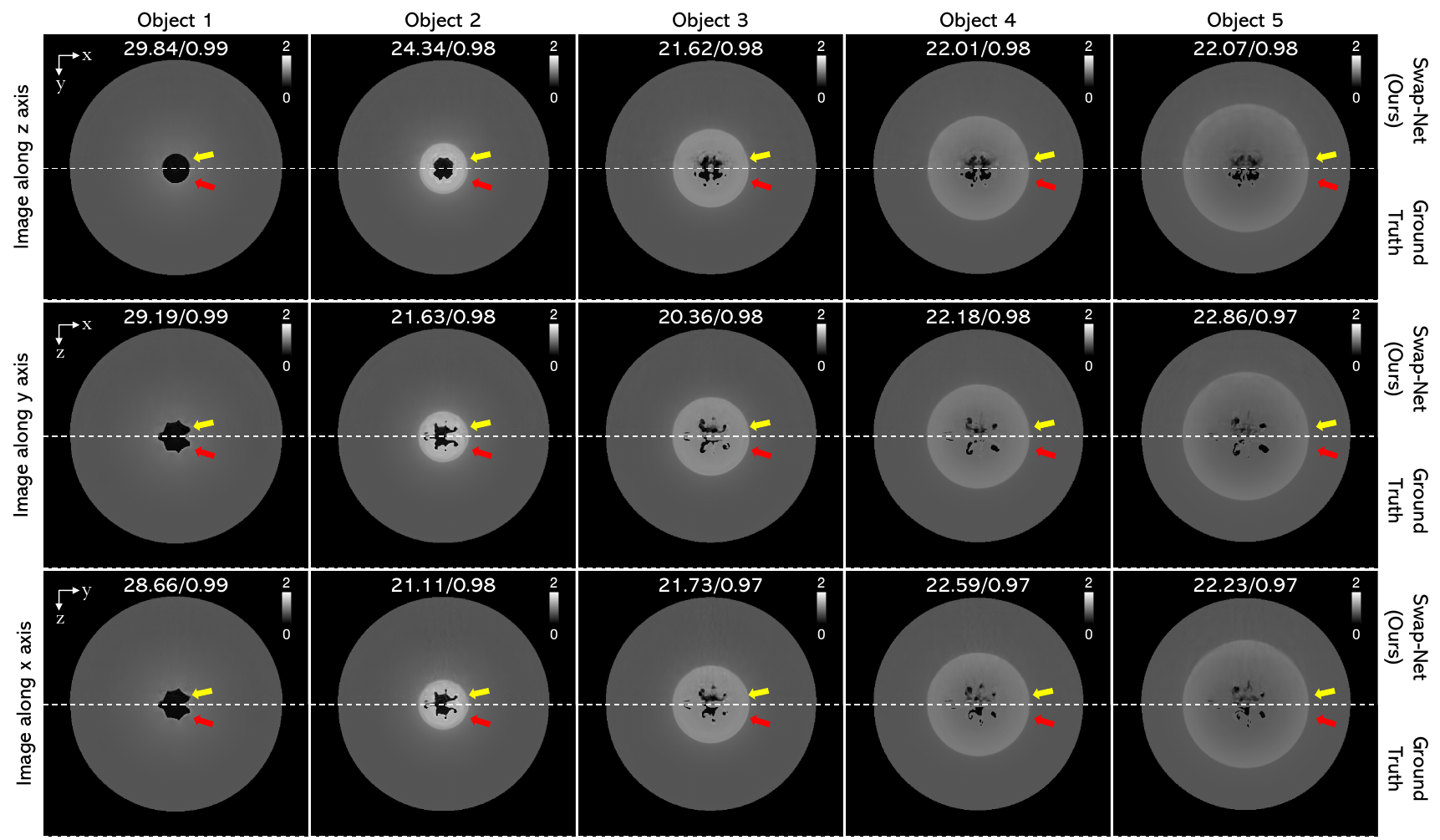}
		\caption{Visual evaluation of Swap-Net on different exemplar ICF double shell test simulations with $4$ projection views
			under AWGN corruption.
			Each row shows different slices of the whole 3D object along $z$, $y$ and $x$ axes, respectively.
			In each row, the images above the dashed line are the images reconstructed from Swap-Net,
			while the images below are the ground truth.
			The top-middle part of each image provides the SNR and SSIM values with respect to the ground truth.
			Arrows in the plots highlight sharp edge regions that are well reconstructed using Swap-Net.
			Swap-Net had
			consistently good performance
			across different 3D objects.}
		\label{fig:compare_objects_awgn}
	\end{figure*}
	
	Fig.~\ref{fig:compare_slices_awgn}
	further illustrates the performance of Swap-Net across different slices in a whole 3D object volume.
	For each slice, we show the side-to-side (top versus bottom) comparison
	between the results of Swap-Net and the corresponding ground truth.
	Using only 4 projection views,
	Swap-Net successfully reconstructed not only the sharp edges but also central details,
	matching well with the ground truth.
	The consistent success of Swap-Net on different slices
	suggests that it can work across the 3D volume,
	highlighting its effectiveness and adaptability.
	Fig.~\ref{fig:compare_objects_awgn}
	additionally shows the visual performance of Swap-Net for different objects.

	\subsection{Additional Study}
\begin{figure*}
		\includegraphics[width=\textwidth]{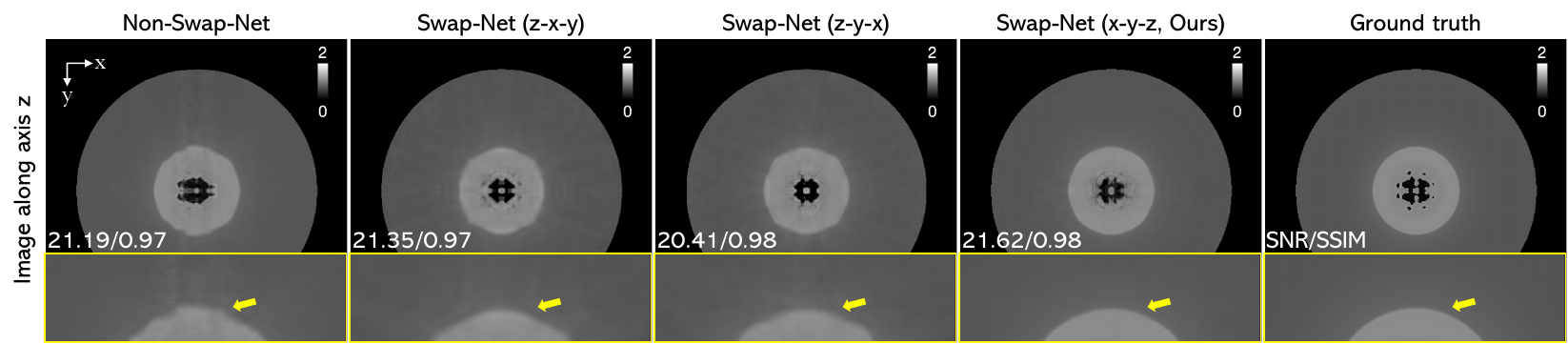}
		\caption{Quantitative and visual evaluation of Swap-Net variants
			with different axes-swapping settings on an exemplar ICF double shell test simulation
			with $4$ projection views
			under AWGN corruption.
			The middle slice of the whole 3D object along $z$ axis is plotted.
			The bottom-left corner of each image provides the SNR and SSIM values,
			and $2\times$ zoomed-in region.
			Arrows in the zoomed-in plots highlight sharp edges that are well reconstructed using Swap-Net
			with swapping order $x$-$y$-$z$.
			Note the the improvement from non-Swap-Net to Swap-Net variants,
			and the the influence of the order of axes-swapping operations in the reconstruction.}
		\label{fig:compare_swap_z_awgn}
	\end{figure*}

	To highlight the contribution of Swap-Net's axes-swapping operation,
	we performed an additional study to examine its influence.
	First, we investigated a \emph{Non-Swap-Net} network identical to the Swap-Net
	but without the axes-swapping operations.
	Comparing to Non-Swap-Net helps to illustrate improvements due to axes-swapping operations.
	We also tested Swap-Net with different axes-swapping orders.
	Letting $z$ denote the CBCT rotation axis,
	we checked the the following axes-swapping orders:
	(a) $z$-$x$-$y$,
	(b) $x$-$z$-$y$,
	and (c) $x$-$y$-$z$,
	namely putting the convolution in the  $xy$ plane
	in the beginning, middle, and end of the Swap-Net pipeline.
	Order (c) $x$-$y$-$z$ is the strategy adopted in our paper.
	Fig.~\ref{fig:compare_swap_z_awgn} presents the reconstruction performance
	of those Swap-Net variants;
	it shows $z$-axis slices
	(similar results were observed for images along $x$ and $y$ axes and therefore were omitted here).
	Clearly, Non-Swap-Net gave the worst results with obvious artifacts,
	and Swap-Net with axes-swapping orders of $z$-$x$-$y$ and $x$-$z$-$y$ did not perform as well as the $x$-$y$-$z$ order.
	We hypothesize this is because the relative worse FBP reconstruction along the $z$ axis
	makes the learning along $z$ more challenging,
	so putting the convolution along $z$ at the end of the Swap-Net pipeline
	allows it to exploit the intermediate object reconstruction.

	\section{Conclusion}
	\label{sec:conclusion}
	
	This paper presents a memory-efficient 2.5D network, namely Swap-Net,
	for handling 3D image reconstruction problems like sparse-view CBCT.
	The major challenge in this problem is
	to reconstruct high-quality 3D images efficiently and accurately
	when only a limited number of projections and training data are available, and when complicated corruptions are presented.
	Swap-Net uses a novel axes-swapping operation
	that allows for sequential convolution along all three dimension of a 3D object.
	We optimized the network weights by minimizing the loss
	between the output of the Swap-Net and the ground-truth 3D images on the training dataset
	using FBP reconstruction as inputs.
	We demonstrated the enhanced performance of our method on sparse-view 3D CBCT image reconstruction
	relative to model-based regularization (such as TV), 2D, and 3D CNNs
	under both AWGN and \todo{non-ideal
		physics including blur and scatter and photon noise} corruptions.
	Our extensive validation elaborated the potential of Swap-Net on producing high-quality images
	from artifact-corrupted measurements.
	Although this paper focuses on CBCT reconstruction,
	our network can be extended to other 3D imaging applications.
	
	In conclusion, our method exploits the lower computational cost of 2D convolution
	while bridging the gap to 3D convolution via axes-swapping operations,
	thereby offering a computationally efficient strategy
	for handling memory-demanding 3D reconstructions.
	Further improvement may be possible by increasing the depth of each convolution block in Swap-Net.
	This work kept the channel dimension to be the same value for all Swap-Net blocks for simplicity;
	future work could optimize the feature dimensions.
	It could also be interesting to explore connections between Swap-Net
	and tensor decomposition methods, e.g.,
	\cite{kilmer2021tensor}.
	Applying Swap-Net to other imaging tasks is planned in the near future.

	\section{Acknowledgement}
	\label{sec:acknowledgement}
	Acknowledgment: This work was supported by the U.S. Department of Energy
	through Los Alamos National Laboratory (LANL).
	The authors acknowledge Jennifer Schei Disterhaupt from LANL
	for contributing to the code for the noise and  scatter model.
	The authors would also like to thank Robert Reinovsky (LANL) for his support of the project.
	
	Data availability statement:
	Data underlying the results presented in this paper are not publicly available at this time
	but may be obtained from the authors upon reasonable request.
	
	Disclosures statement: The authors declare no conflicts of interest.
	
	%
	
	\bibliographystyle{IEEEtran}
	\bibliography{ref}
\end{document}